\documentstyle[epsfig,preprint,aps]{revtex}
\newcommand{\ba}{\begin{eqnarray}}
\newcommand{\ea}{\end{eqnarray}}
\newcommand{\be}{\begin{equation}}
\newcommand{\ee}{\end{equation}}
\newcommand{\bd}{\begin{displaymath}}
\newcommand{\ed}{\end{displaymath}}

\preprint{\normalsize NUC-MINN-03/4-T}
\title{Rate Equation Network for Baryon Production in High Energy Nuclear 
Collisions}

\author{Pasi Huovinen$^{1,2,3}$ and Joseph Kapusta$^{1}$}
\address{$^{1}$School of Physics and Astronomy, University of Minnesota\\
         Minneapolis, Minnesota 55455, USA\\
         $^{2}$Helsinki Institute of Physics, P.O.\ Box 64\\
         FIN-00014 University of Helsinki, Finland\\
         $^{3}$Department of Physics, P.O.\ Box 35\\
         FIN-40014 University of Jyv\"askyl\"a, Finland}

\draft

\begin{document}

\maketitle

\begin{abstract}
We develop and solve a network of rate equations for the production of baryons 
and anti-baryons in high energy nuclear collisions.  We include all members of 
the baryon octet and decuplet and allow for transformations among them.  This 
network is solved during a relativistic 2+1 hydrodynamical expansion of the 
of the hot matter created in the collision.  As an application we compare to the 
number of protons, lambdas, negative cascades, and omega baryons measured at 
mid-rapidity in central collisions of gold nuclei at 65 GeV per nucleon at the 
Relativistic Heavy Ion Collider (RHIC).
\end{abstract}
\vspace{0.25in}

\pacs{PACS: 12.38.Mh, 25.75.Dw, 25.75.-q}

\section{Introduction}

It has been a challenge to understand the relatively high abundances of baryons 
and anti-baryons produced in high energy gold-gold collisions at the Brookhaven 
National Laboratory RHIC (Relativistic Heavy Ion
Collider).  The center of mass collision energies have ranged between 56 and 200 
GeV per nucleon pair.  This energy is so high that baryon/anti-baryon pairs can 
readily be created.  To very good accuracy the relative abundances, including 
multiply strange hyperons, are consistent with them being in chemical 
equilibrium at a temperature of $170 \pm 10$ MeV and a baryon chemical potential 
on the order of tens of MeV \cite{chem1,chem2}.  Furthermore, the ratio of 
baryons to mesons grows with increasing transverse momentum, reaching one at 
$p_T \approx 2$ GeV/c \cite{phenix-pions}.  
A frequently given explanation of the first result is that the hadrons are 
created in chemical equilibrium at some temperature and chemical potential, 
possibly at the end of a quark-gluon to hadron phase transition.  In the absence 
of detailed information on the dynamics it is quite reasonable to postulate that 
phase space is filled randomly.   For a big system, this is equivalent 
mathematically to momentary or instantaneous thermal and chemical equilibrium.  
A possible explanation of the second result is that since the transverse 
momentum is proportional to mass, collective fluid-like flow velocity of the 
expanding matter would boost heavier particles to higher transverse momentum.  A 
nice overview of these results can be obtained from the Proceedings of Quark 
Matter 2002 \cite{QM2002}.

The results on baryon production at RHIC are exciting and stimulating.  A 
parametrization of them in terms of temperature $T$, baryon chemical potential 
$\mu$ arising from the initial baryon number of the gold nuclei, and transverse 
flow velocity $v_T$ is a first step.  The next step is to understand them in 
terms of microscopic, dynamical processes.  In this paper we set up a network of 
rate equations to describe the production of baryon/anti-baryon pairs (such as 
$p\bar{\Lambda}$ and $\Sigma\bar{\Delta}$) including all members of the octet 
and decuplet.  We also include transformations among them (such as $\Delta^+ 
\leftrightarrow p\pi^0$).  The rate of production of octet-octet pairs was 
derived by Kapusta and Shovkovy \cite{KS}.  To compare with experimental data it 
is important to include the production of octet-decuplet and decuplet-decuplet 
pairs too.  Unfortunately there is insufficient experimental information to 
determine these as accurately.  Therefore, we use a simple form of SU(3) flavor 
symmetry to estimate these rates.  Then we solve the network of rate equations 
in a 2+1 dimensional hydrodynamical model of the expanding high energy density 
matter in heavy ion collisions which has been tuned to reproduce the measured 
pion multiplicity~\cite{phenix-pions}.  Since this is a set of coupled
differential equations initial conditions are required.  We begin to solve the 
rate equations in the mixed phase of a first order quark-gluon to
hadron phase transition with three values of $T_c$, namely, 165, 180,
and 200 MeV.  In one scenario we assume that all baryons are in
chemical equilibrium initially, and in the other we assume no baryons
initially.  Other scenarios are certainly possible.  We plot the
resulting abundances as a function of the local (freeze-out)
temperature $T_f$ and compare to the measured numbers of $p$,
$\Lambda$, $\Xi^-$, and $\Omega$ in central gold-gold collisions at
mid-rapidity at a beam energy of 65 GeV per nucleon at RHIC.  We also
study what happens when the octet-decuplet and decuplet-decuplet rates
are increased or decreased by a factor of two.  Comparison to the
transverse momentum distributions of these baryons will be done in a
later paper.

\section{Network of rate equations}

We need to specify the dynamical variables, measure or calculate the
microscopic rates, and then solve the resulting network of rate
equations.  We shall do each of these in turn.

Various members of the baryon (or anti-baryon) octet and decuplet have
been or will be measured in central heavy ion collisions at the CERN
Super Proton Synchrotron (SPS) and at RHIC.  Therefore the variables
will be all members of those multiplets.  There are many mesons that
are much lighter than the baryons, such as the pion and the $\rho$ and
$\omega$ vector mesons to name just a few.  The baryons are also much
heavier than the three lightest species of quarks when the latter are
given their current quark masses.  Therefore it ought to be a good
approximation to consider the mesons and/or the quarks and gluons as
providing a thermal bath in which baryon/anti-baryon pairs can be
created or destroyed.  Since kinetic equilibrium is usually reached
much quicker than chemical equilibrium it is reasonable to take the
kinetic energy distribution of the baryons and anti-baryons as
approximately thermal.  We assume in this paper that only the absolute
number of each species deviates from local chemical equilibrium.
These approximations can be relaxed but, as they are, the calculations
become increasingly complex.  Ultimately one would reach the point
where microscopic cascade calculations would be necessary, and then
one encounters the problem of multi-particle initial states in the
quasi-localized interactions.  That problem is largely avoided in the
present approximation because of the thermal averaging of the
microscopic rates (see below).

Under the above conditions, the rate equation for the spatial density of a 
baryon species $b$ is
\ba
\frac{dn_b}{dt} &=& \sum_{\bar{b}'} R(b\bar{b}')
\left[ 1 - \frac{n_{\bar{b}'} n_b}{n_{\bar{b}'}^{\rm equil} n_b^{\rm equil}}
\right] - \frac{n_b}{V} \frac{dV}{dt} \nonumber \\
& & \mbox{} + \sum_{b'} \Gamma(b' \rightarrow b+X)
\left[ n_{b'} - \frac{n_{b'}^{\rm equil} \, n_b}
{n_b^{\rm equil} } \right] - \sum_{b'} \Gamma(b \rightarrow b'+X)
\left[ n_{b} - \frac{n_{b}^{\rm equil} \, n_{b'}}
{n_{b'}^{\rm equil} } \right] 
\ea
The first term on the right involves the rate for producing the specified 
baryon/anti-baryon pair by strong interaction currents, $R(b\bar{b}') =$ the 
number of such pairs produced per unit volume per unit time, and is the
driving term in the rate equation.  Of course, the same pair can annihilate
each other, and this is related to the production rate by detailed balance.
The factor in square brackets enforces detailed balance.  The
$n_{b'}^{\rm equil}$ is the equilibrium density of the species $b'$ as
represented by the temperature and chemical potentials at the time $t$.
The second term is a dilution term; the density will decrease in inverse
proportion to the volume for an expanding system.  The third term arises if
there exists a baryon species that can decay via the strong interactions into
the baryon species of interest.  The quantity in square brackets following it
allows for the inverse reaction and satisfies detailed balance. The last term
arises if the baryon species of interest can decay into another baryon via
the strong interactions.  There is one first order in time, nonlinear, rate
equation for each species of baryon.  This makes a coupled network of
differential equations.

The decay rate $\Gamma(b' \rightarrow b+X)$ is just the inverse
lifetime for the specified decay, namely, $1/\tau(b' \rightarrow
b+X)$.  These are taken from the Particle Data Tables \cite{PDG}.  
We include the following decays: $\Delta \rightarrow N+\pi$,
$\Sigma^* \rightarrow \Lambda+\pi$, $\Sigma^* \rightarrow \Sigma+\pi$,
$\Xi^* \rightarrow \Xi+\pi$.

The rate $R$ is the crucial ingredient in this network of rate
equations. It was derived in quite some detail by Kapusta and Shovkovy
\cite{KS} for the case of baryons and anti-baryons in the lowest
octet.  The basis for that calculation involved an effective
current-current interaction between the baryons and strong interaction
currents such as the vector and axial-vector.  A combination of hadron
phenomenology, experimental data on cross sections, and SU(3) flavor
symmetry allows the rates to be expressed in terms of the spectral
densities of the strong interaction currents.  Since the $\sqrt{s}$ is
so large, with a minimum value determined by threshold production of
the baryon/anti-baryon pair, these spectral densities can be taken
from perturbative QCD.  For the production of baryons in the octet,
the rate was expressed as \cite{KS}
\be
R_{8,8}(b_1\bar{b}_2) = C_+(b_1\bar{b}_2) {\cal R}_+(b_1\bar{b}_2) +
C_-(b_1\bar{b}_2) {\cal R}_-(b_1\bar{b}_2) \, .
\ee
Here
\ba
{\cal R}_{\pm}(b_1\bar{b}_2) &=&
\frac{9(1+\alpha_s/\pi)T^8}{4(2\pi)^5f_{\pi}^4} z_1^2 z_2^2
\left\{ 4z_1K_1(z_1)K_2(z_2) + 4z_2K_1(z_2)K_2(z_1) \right. \nonumber \\
&& \left. \pm (z_1 \pm z_2)^2 K_1(z_1)K_1(z_2) +
[16+(z_1\pm z_2)^2] K_2(z_1)K_2(z_2) \right\}
F_{\rm ANN}^2(\bar{s})
\ea
where $z_i=m_i/T$, $T$ is the temperature, and $F_{\rm ANN}^2(s)$ is an 
annihilation form factor evaluated at the average value
$\bar{s} = (m_1+m_2)^2 + 3(m_1+m_2)T$. A simple parametrization of this
form factor which is consistent with nucleon/anti-nucleon annihilation
data was found to be
\be
F_{\rm ANN}(s) = \frac{1}{2.21 + [s-(m_1+m_2)^2]/\Lambda^2} \, ,
\ee
with $\Lambda = 1.63$ GeV.  According to the latest data analysis \cite{PDG}
the strong interaction coupling is $\alpha_s(m_{\tau}^2) = 0.35 \pm 0.03$.
The $\pm$ correspond to vector/axial-vector contributions to the rate.  The 
numerical coefficients $C_{\pm}(b_1\bar{b}_2)$ are given in \cite{KS};
they are all of order one.

There is unlikely ever to be sufficient experimental information from
hadron-hadron scattering to pin down the production rates for baryons in the 
decuplet since they are all unstable with lifetimes less than $10^{-10}$ 
seconds.  Therefore, we have estimated the rates and parametrize them as 
follows.
\ba
R_{8,10}(b_1\bar{b}_2) &=& C_{8,10}(b_1\bar{b}_2) {\cal R}(b_1\bar{b}_2)
\nonumber \\
R_{10,10}(b_1\bar{b}_2) &=& C_{10,10}(b_1\bar{b}_2) {\cal R}(b_1\bar{b}_2)
\ea
We take the coefficients of the vector and axial-vector contributions to be 
equal which means that
\ba
{\cal R}(b_1\bar{b}_2) &=& 
{\cal R}_+(b_1\bar{b}_2) + {\cal R}_-(b_1\bar{b}_2) \nonumber \\
&=& \frac{9(1+\alpha_s/\pi)T^8}{2(2\pi)^5f_{\pi}^4} z_1^2 z_2^2
\left\{ 4z_1K_1(z_1)K_2(z_2) + 4z_2K_1(z_2)K_2(z_1) \right. \nonumber \\
& & \mbox{} \left. 
+ 2z_1z_2 K_1(z_1) K_1(z_2)
+ [16+z_1^2 + z_2^2] K_2(z_1)K_2(z_2) \right\}
F_{\rm ANN}^2(\bar{s}) \, .
\ea 
The coefficients $C_{8,10}(b_1\bar{b}_2)$ and
$C_{10,10}(b_1\bar{b}_2)$ are displayed in Tables I and II,
respectively.  We have constructed them as follows.  First, an entry
in the table is zero if the quantum numbers of the pair do not match
the quantum numbers of the vector or axial-vector meson nonets.
Second, exact isospin symmetry within SU(2) multiplets is used.
Third, the sum of the coefficients of each column of Table I equals 4,
which is the order of magnitude of the corresponding octet-octet
coefficients $C_{8,8}$; the sum of the coefficients of each row Table
I is either $4\frac{2}{3}$, 5, or $5\frac{1}{2}$.  Flavor SU(3)
symmetry cannot give more information than this.  Similarly the sum of
the coefficients of each column (and row) of Table II equals 6.  In
our numerical work we shall study the consequences of increasing or
decreasing these coefficients overall by a factor of 2.

There will always be more baryons than anti-baryons present in the
final state of a heavy ion collision because the colliding nuclei have
a net baryon number.  However, this asymmetry is reduced as the beam
energy increases because produced baryons always come with an
anti-baryon due to baryon number conservation.  The pair production is
an increasing function of beam energy and so the initial number of
baryons from the nuclei becomes a smaller and smaller fraction of the
total.  At the highest energies at RHIC the baryon chemical potential
is usually estimated to be on the order of 10 MeV when the temperature
is 150 MeV.  For simplicity of calculation and presentation we shall
take the baryon chemical potential to be exactly zero in this paper.
To remove the difference between the experimental data and our
calculation due to this approximation, we compare our results to the
observed average yield of a baryon and antibaryon, for example $(p+\bar{p})/2$,
not to the observed baryon yield as such.

\section{Comparison to data}

Baryon and anti-baryon production cannot begin until the local energy
density is low enough, that is, when the matter is in the hadronic
phase as opposed to the quark-gluon phase.  This will occur during the
expansion stage of a high energy nucleus-nucleus collision.  This
stage is frequently modeled with hydrodynamics~\cite{hydrorev}.  We
shall do so in this paper too.  We will use a 2+1 dimensional
description of the final stage expansion that takes into account
transverse expansion.  The details of these calculations have been
given many times before and so we just refer the reader to those
papers~\cite{hydro}.  Briefly, the expansion begins in the quark-gluon
plasma phase with an energy density adjusted to reproduce the measured
pion rapidity density at mid-rapidity.  The phase transition is first
order; we have chosen $T_c$ to be 165, 180, and 200 MeV and test the
sensitivity of baryon production to it.

The network of rate equations is solved within each co-moving cell in coordinate
space.  Different cells evolve somewhat differently in space and time, and this
makes for a very computationally intense task.  At each point in time, as 
measured by an observer at rest in the center-of-momentum frame of the central 
gold-gold collision, each cell has its own temperature.  This local temperature 
enters in the production rates and in the local chemical equilibrium densities. 
In order to concisely display the results of our calculations we plot the 
solutions to the network of rate equations as a function of the local 
temperature and not as a function of the local time. 
Because each fluid cell has its own temperature at any given time, we
must integrate over a constant temperature hypersurface to get the baryon
rapidity density at this temperature. Thus the value at a fixed
temperature does not correspond to any particular time.
This choice is rather natural since baryon production begins
at a system-wide value of the temperature equal to $T_c$ and ends at a supposed
freeze-out temperature $T_f$ where the hadrons lose local thermal equilibrium 
and begin their free-streaming stage.

In figure 1 we display the results of numerical solution of the rate equation 
network for $p$, $\Lambda$, $\Xi^-$ and $\Omega$.  Plotted is the ratio of the 
calculated density to the equilibrium density at that particular temperature.  
No weak decays of unstable baryons have been allowed for at this point.  The 
rate equations require an initial condition and we have chosen two: all baryons 
in chemical equilibrium at $T_c$, and no baryons present at all at $T_c$.  The 
former are represented in the figure by the upper set of thick curves 
while the latter is represented by the lower set of thin curves. Obviously this 
calculation has used $T_c$ = 180 MeV, but it is representative of other choices. 
When baryons are present initially with their equilibrium abundances, they 
evolve in such a way that they always stay above the equilibrium abundance at 
each temperature below $T_c$.  The reason is that the system expands more 
rapidly than chemical reactions can keep up with.  In other words, the typical 
annihilation rates are smaller than the expansion rate.  When no baryons are 
present initially, their abundance at first builds up rapidly.  The reason that 
there is a finite abundance at $T_c$ already is that the cells generally remain 
within the mixed phase at $T_c$ for a finite time span, thus allowing a buildup 
before the temperature drops below $T_c$.  In the range from 120 to 130 MeV, 
these baryon species have caught up to the equilibrium abundance at that 
temperature. Thereafter they are above the equilibrium values for the same 
reason as stated above, namely, that annihilation rates are generally smaller 
than the expansion rate.  Naturally, the abundances with no baryons present 
initially never catch up with the abundances where baryons were initially 
created in chemical equilibrium.

In figures 2-5 we show the rapidity density at mid-rapidity as a function of
freeze-out temperature for $p$, $\Lambda$, $\Xi^-$ and $\Omega$, respectively.
The upper set of curves result from having baryons in chemical equilibrium
initially, while the lower set of curves result from having no baryons present
initially.  The central curve in each set results from using the standard set of
octet-decuplet and decuplet-decuplet coefficients shown in the Tables. The upper
and lower curves in the lower set result from increasing and decreasing these
coefficients by a factor of 2 respectively (the octet-octet coefficients are
left unchanged). In the upper set of curves the effect is reversed: the larger
rates lead to smaller yields.  Also shown in these figures are the experimental 
data from RHIC experiments. Because we use the approximation of zero baryon 
chemical potential, we compare our result to
the measured average yield of baryon and antibaryon. The $(p+\bar{p})/2$ and 
$(\Lambda+\bar{\Lambda})/2$ data come from PHENIX \cite{PHENIX} and the 
$(\Xi^-+\bar{\Xi}^+)/2$ and $(\Omega+\bar\Omega)/2$ data 
come from STAR \cite{STAR}. The calculation and the PHENIX data are for 5\% most 
central collisions whereas the STAR data are for 10\% most central collisions.
By comparing the hydrodynamically calculated pion yields at different 
centralities we have estimated that this leads to about $10$\% larger yield in 
our calculation compared to the data. The darker central band in each figure 
represents statistical errors only while the lighter outer band includes 
systematic errors as well.  (For $\Omega$ the systematic error is smaller than 
the statistical error.)  We have performed decays of unstable baryons 
appropriate to the particular measurement.  For example, protons coming from the 
weak decay $\Lambda \rightarrow p+\pi^-$ are not included in figure 2, but 
protons from the decay $\Sigma^+ \to p+\pi^0$ are. 

What we learn from this set of figures is that the baryon production rates are 
too small in comparison to the expansion rate to reproduce any of the 
experimental data when there are no baryons present initially.  There must be a 
significant abundance of baryons present already at the beginning of the 
hadronic phase.  Assuming the baryons to be produced in chemical equilibrium and 
then evolving them during the expansion of the matter provides agreement with 
the data on $p$, $\Lambda$, and $\Xi^-$ within systematic error bars for $120 
\leq T_f \leq T_c$.  However, the theoretical calculations produce about 50\% 
more $\Omega$ than observed.  It is interesting that this magnitude of 
discrepancy does not occur for the $\Xi^-$, even though it has two valence 
s-quarks.  Oftentimes in purely statistical models an s-quark fugacity factor is 
used to fit the data.  A typical value might be 0.9, so that the $\Lambda$ yield 
would be multiplied by 0.9, the $\Xi^-$ yield by $(0.9)^2$, and the $\Omega$ 
yield by $(0.9)^3$.  Any such suppression factor should be an outcome of the 
results of a rate equation calculation like the present one, and cannot be done 
arbitrarily at the end of the calculation to fit data. On the other hand, if one 
had confidence that there was a first-order phase transition with this numerical 
value of $T_c$ and that the expansion was described adequately by hydrodynamics, 
then one could adjust the initial conditions to match the experimental data.

Next we test the sensitivity of the results to the critical temperature.  The 
results are shown in figures 6-9.  In these calculations we have used the set of 
coefficients from Tables I and II.  The calculated abundances are generally 
increasing functions of $T_c$, chosen here to be 165, 180 and 200 MeV.  As seen 
in figures 2-5, the abundances calculated when no baryons are present initially 
are all below the data.  When the baryons are in chemical equilibrium initially, 
the data on $\Lambda$ and $\Xi^-$ are bracketed by the curves.  The $p$ are 
consistent, within systematic errors, with 180 and 200 MeV.  The $\Omega$ data 
is consistent with 165 MeV but is below the 180 and 200 MeV curves.  However, 
there is no value of $T_c$ which provides agreement between the calculated 
abundance and the observed one for all four species of baryons.
 
\section{Conclusion}

In this paper we have estimated the thermal production rates of
baryon/anti-baryon pairs when one or both of them are members of the spin-3/2 
decuplet.  This estimate is based on the thermal production rates when both are 
members of the spin-1/2 octet \cite{KS}.  The latter rates are quite
well-founded in hadronic phenomenology and QCD, whereas the rates involving the 
spin-3/2 baryons are somewhat uncertain as they are unstable and beams cannot be 
made to measure cross sections.  We used exact isospin symmetry within 
multiplets; flavor SU(3) was used too, but it is insufficient to pin down all 
the rate coefficients.  We made numerical estimates based on octet-octet 
production and ultimately tested the sensitivity of the results to changes of a 
factor 2 in the coefficients.

We used 2+1 dimensional hydrodynamics, which includes transverse expansion, to 
model the expansion stage of a central gold-gold collision at RHIC.  This 
provided the thermal bath within which the network of rate equations was solved. 
The network of equations requires 
initial conditions.  We chose two interesting limits to study in detail: no 
baryons present initially and baryons initially in chemical equilibrium.  We 
also studied the sensitivity of the results to the critical temperature of an 
assumed first order phase transition from quarks and gluons to hadrons.

When comparison is made to RHIC data on $p$, $\Lambda$, $\Xi^-$ and $\Omega$, 
the most important conclusion is that the calculated production rates are too 
small to compensate for the rapid expansion of the matter as described by 
hydrodynamics if there are no baryons present at the end of the phase 
transition.  There is just not enough time to build up the abundances of these 
baryons starting from nothing.  On the other hand, the observed abundances of 
$p$, $\Lambda$, and $\Xi^-$ can be reproduced when this network of rate 
equations is solved within a hydrodynamic expansion if baryons are taken to be 
in chemical equilibrium when each fluid cell is converted from quarks and gluons 
to hadrons.  The observed abundance of $\Omega$, however, is generally less than 
the calculated one, and when the parameters are chosen to obtain the correct 
$\Omega$ abundance the other baryons are underproduced.  So the $\Omega$ remains 
a bit of a puzzle.

The results obtained here are not trivial nor are they obvious.  The interplay 
among the initial conditions, the expansion dynamics (in particular the 
expansion rate), and the production rates of octet and decuplet baryons is 
intricate and subtle.  If the baryons are produced in chemical equilibrium at 
$T_c$, their numbers will decrease as the system expands and cools by an amount 
that depends on the annihilation rates compared to the expansion rate.  If no 
baryons are present at $T_c$, their numbers will at first increase with time 
until local equilibrium is achieved at some temperature below $T_c$ which is 
generally different for each species.  After that there will be an overabundance 
of baryons relative to equilibrium, and their numbers will begin to decrease.

Among the extensions of this project under investigation are different choices 
of the initial abundances and the transverse momentum distributions of the 
various baryon species.  The latter would allow us to study the expansion 
dynamics in finer details than just the overall dN/dy at mid-rapidity.   

\section*{Acknowledgements}

This work was supported by the US Department of Energy under grant
DE-FG02-87ER40328.

\newpage

\noindent
Table I: Relative strengths of coefficients in the\\
expressions for the rates.\\[5mm]
\begin{tabular}{|c||c|c|c|c|c|c|c|c|c|c|}
\hline
\hspace*{10mm}& $~\Delta^{++}~$  & $~\Delta^+~$ & $~\Delta^0~$ & $~\Delta^-~$ & 
$~\Sigma^{*+}~$ & $~\Sigma^{*0}~$ &
$~\Sigma^{*-}~$ & $~\Xi^{*0}~$ & $~\Xi^{*-}~$ & $~\Omega~$\\
\hline
\hline
%%%%%%%%%%%%%%%%%%%%%%%%%%%%%%%%%%%%%%%%%%%%%%%%
$\overline{p}$ &
$2$ & $4/3$ & $2/3$ & $0$ & $1$ & $1/2$ & $0$ & $0$ & $0$ & $0$\\
\hline
%%%%%%%%%%%%%%%%%%%%%%%%%%%%%%%%%%%%%%%%%%%%%%%%
$\overline{n}$ &
$0$ & $2/3$ & $4/3$ & $2$ & $0$ & $1/2$ & $1$ & $0$ & $0$ & $0$\\
\hline
%%%%%%%%%%%%%%%%%%%%%%%%%%%%%%%%%%%%%%%%%%%%%%%%
$\overline{\Lambda}$ &
$0$ & $0$ & $0$ & $0$ & $1$ & $1$ & $1$ & $1$ & $1$ & $0$\\
\hline
%%%%%%%%%%%%%%%%%%%%%%%%%%%%%%%%%%%%%%%%%%%%%%%%
$\overline{\Sigma}^{+}$  &
$2$ & $2/3$ & $0$ & $0$ & $1/2$ & $1/2$ & $0$ & $1$ & $0$ & $0$\\
\hline
%%%%%%%%%%%%%%%%%%%%%%%%%%%%%%%%%%%%%%%%%%%%%%%%
$\overline{\Sigma}^{0}$  &
$0$ & $4/3$ & $4/3$ & $0$ & $1/2$ & $0$ & $1/2$ & $1/2$ & $1/2$ & $0$\\
\hline
%%%%%%%%%%%%%%%%%%%%%%%%%%%%%%%%%%%%%%%%%%%%%%%%
$\overline{\Sigma}^{-}$ &
$0$ & $0$ & $2/3$ & $2$ & $0$ & $1/2$ & $1/2$ & $0$ & $1$ & $0$\\
\hline
%%%%%%%%%%%%%%%%%%%%%%%%%%%%%%%%%%%%%%%%%%%%%%%%
   $\overline{\Xi}^{0}$ &
$0$ & $0$ & $0$ & $0$ & $1$ & $1/2$ & $0$ & $1$ & $1/2$ & $2$\\
\hline
%%%%%%%%%%%%%%%%%%%%%%%%%%%%%%%%%%%%%%%%%%%%%%%%
  $\overline{\Xi}^{-}$ &
$0$ & $0$ & $0$ & $0$ & $0$ & $1/2$ & $1$ & $1/2$ & $1$ & $2$\\
\hline
%%%%%%%%%%%%%%%%%%%%%%%%%%%%%%%%%%%%%%%%%%%%%%%%
\end{tabular}
\vspace{15mm}

\newpage

\noindent
Table II: Relative strengths of coefficients in the\\
expressions for the rates.\\[5mm]
\begin{tabular}{|c||c|c|c|c|c|c|c|c|c|c|}
\hline
\hspace*{10mm}& $~\Delta^{++}~$  & $~\Delta^+~$ & $~\Delta^0~$ & $~\Delta^-~$ & 
$~\Sigma^{*+}~$ & $~\Sigma^{*0}~$ &
$~\Sigma^{*-}~$ & $~\Xi^{*0}~$ & $~\Xi^{*-}~$ & $~\Omega~$\\
\hline
\hline
%%%%%%%%%%%%%%%%%%%%%%%%%%%%%%%%%%%%%%%%%%%%%%%%
$\overline{\Delta}^{++}$ &
$12/5$ & $8/5$ & $0$ & $0$ & $2$ & $0$ & $0$ & $0$ & $0$ & $0$\\
\hline
%%%%%%%%%%%%%%%%%%%%%%%%%%%%%%%%%%%%%%%%%%%%%%%%
$\overline{\Delta}^+$ &
$8/5$ & $4/15$ & $32/15$ & $0$ & $2/3$ & $4/3$ & $0$ & $0$ & $0$ & $0$\\
\hline
%%%%%%%%%%%%%%%%%%%%%%%%%%%%%%%%%%%%%%%%%%%%%%%%
$\overline{\Delta}^0$ &
$0$ & $32/15$ & $4/15$ & $8/5$ & $0$ & $4/3$ & $2/3$ & $0$ & $0$ & $0$\\
\hline
%%%%%%%%%%%%%%%%%%%%%%%%%%%%%%%%%%%%%%%%%%%%%%%%
   $\overline{\Delta}^-$  &
$0$ & $0$ & $8/5$ & $12/5$ & $0$ & $0$ & $2$ & $0$ & $0$ & $0$\\
\hline
%%%%%%%%%%%%%%%%%%%%%%%%%%%%%%%%%%%%%%%%%%%%%%%%
   $\overline{\Sigma}^{*+}$  &
$2$ & $2/3$ & $0$ & $0$ & $2/3$ & $2/3$ & $0$ & $2$ & $0$ & $0$\\
\hline
%%%%%%%%%%%%%%%%%%%%%%%%%%%%%%%%%%%%%%%%%%%%%%%%
   $\overline{\Sigma}^{*0}$  &
$0$ & $4/3$ & $4/3$ & $0$ & $2/3$ & $0$ & $2/3$ & $1$ & $1$ & $0$\\
\hline
%%%%%%%%%%%%%%%%%%%%%%%%%%%%%%%%%%%%%%%%%%%%%%%%
   $\overline{\Sigma}^{*-}$ &
$0$ & $0$ & $2/3$ & $2$ & $0$ & $2/3$ & $2/3$ & $0$ & $2$ & $0$\\
\hline
%%%%%%%%%%%%%%%%%%%%%%%%%%%%%%%%%%%%%%%%%%%%%%%%
   $\overline{\Xi}^{*0}$ &
$0$ & $0$ & $0$ & $0$ & $2$ & $1$ & $0$ & $2/3$ & $1/3$ & $2$\\
\hline
%%%%%%%%%%%%%%%%%%%%%%%%%%%%%%%%%%%%%%%%%%%%%%%%
  $\overline{\Xi}^{*-}$ &
$0$ & $0$ & $0$ & $0$ & $0$ & $1$ & $2$ & $1/3$ & $2/3$ & $2$\\
\hline
%%%%%%%%%%%%%%%%%%%%%%%%%%%%%%%%%%%%%%%%%%%%%%%%
$\overline{\Omega}$ &
$0$ & $0$ & $0$ & $0$ & $0$ & $0$ & $0$ & $2$ & $2$ & $2$\\
\hline
%%%%%%%%%%%%%%%%%%%%%%%%%%%%%%%%%%%%%%%%%%%%%%%%
\end{tabular}

\begin{figure}
\epsfig{file=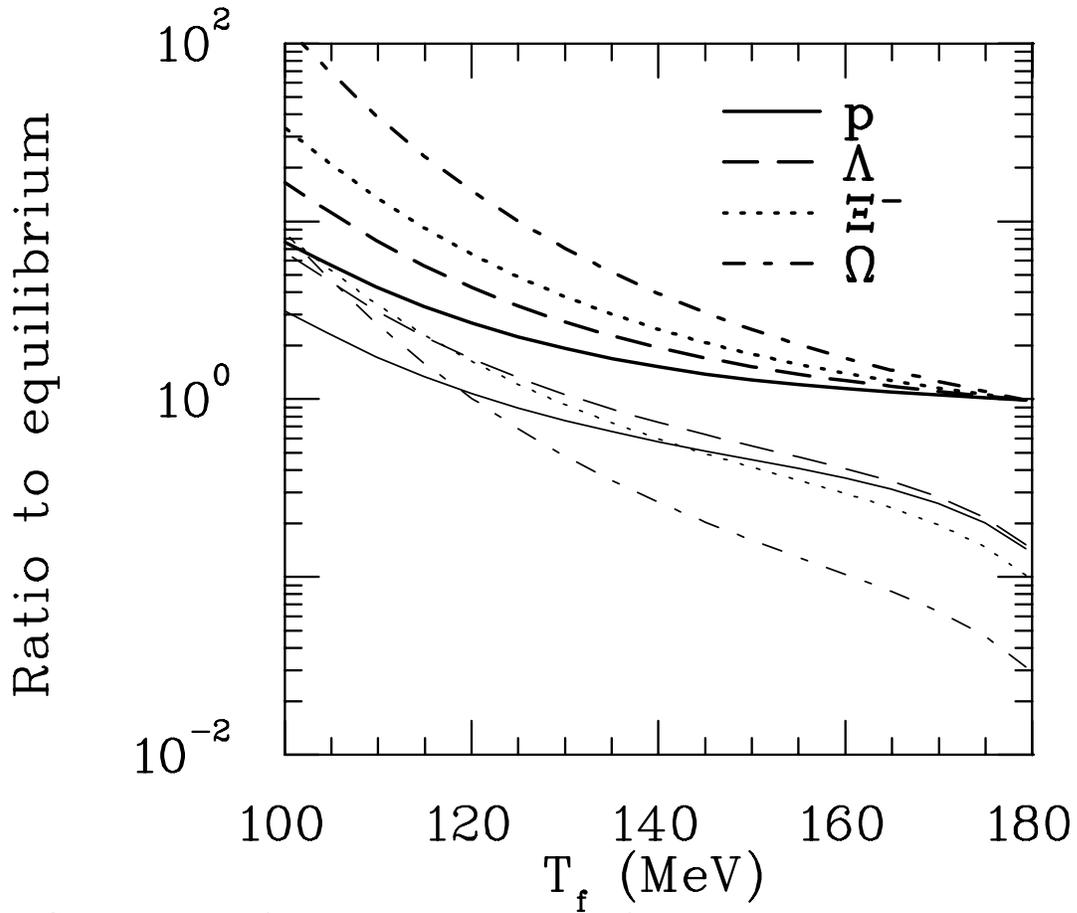,width=14cm}
\caption{The ratio of the calculated abundance of the indicated species
to the chemical equilibrium value as a function of the local temperature.
The upper set of curves start with the baryons in equilibrium at $T_c$,
the lower set start with no baryons.}
\label{ratio_fig}
\end{figure}

\begin{figure}
\epsfig{file=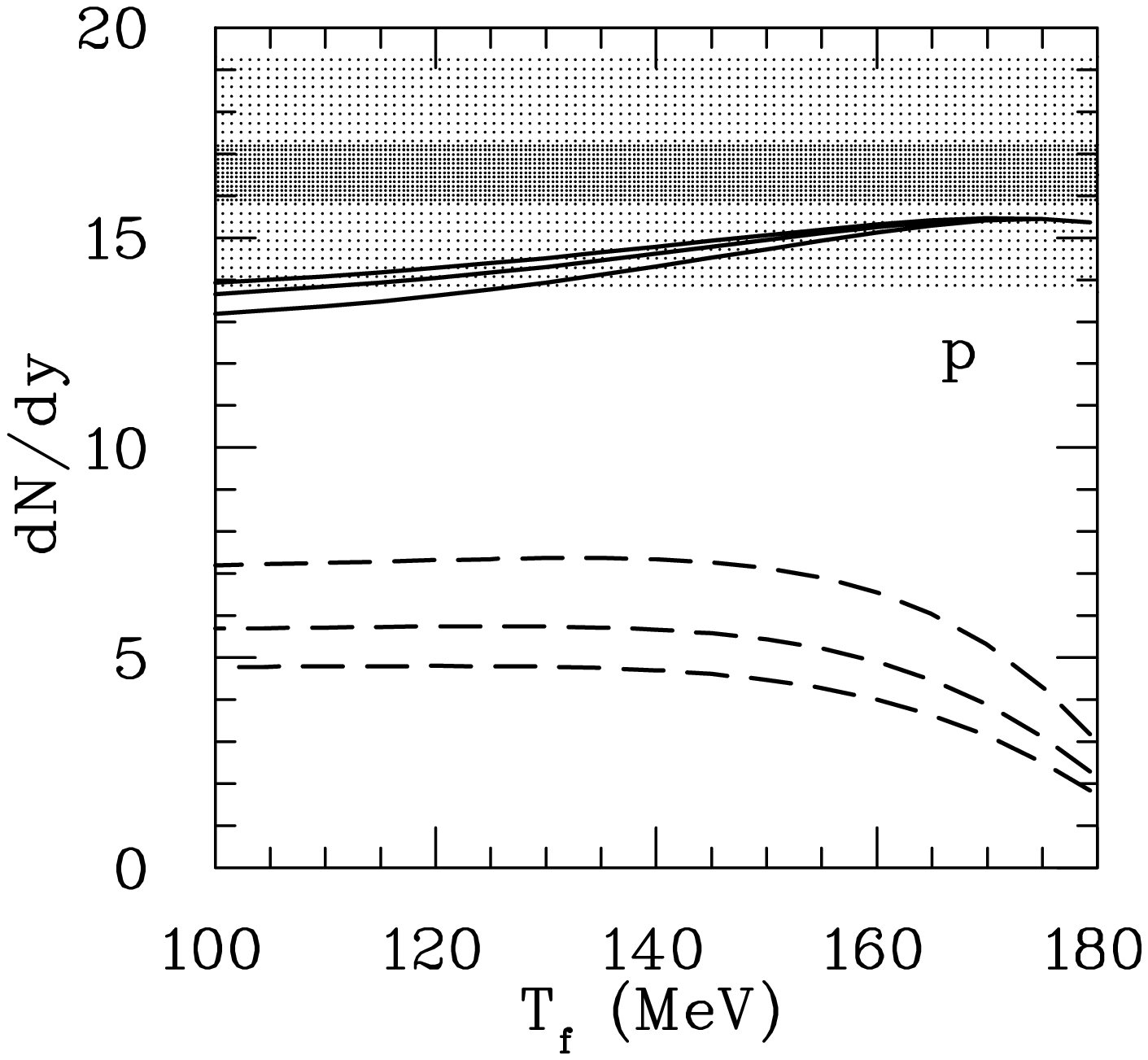,width=14cm}
\caption{The number of protons at central rapidity in 5\% most central
gold-gold collisions as a function of local temperature.  The solid
curves start with baryons in equilibrium, the lower set with no
baryons.  The central curve in each set uses the coupling coefficients
given in the Tables, while the other two use values larger and smaller
by a factor of two. With the exception of $\Lambda$, baryons unstable
to strong or weak decays have been decayed.  The data are from the
PHENIX collaboration~\protect\cite{PHENIX} for the same centrality.
The dark band represents statistical and the light band systematic
errors.}
\label{proton_fig1}
\end{figure}

\begin{figure}
\epsfig{file=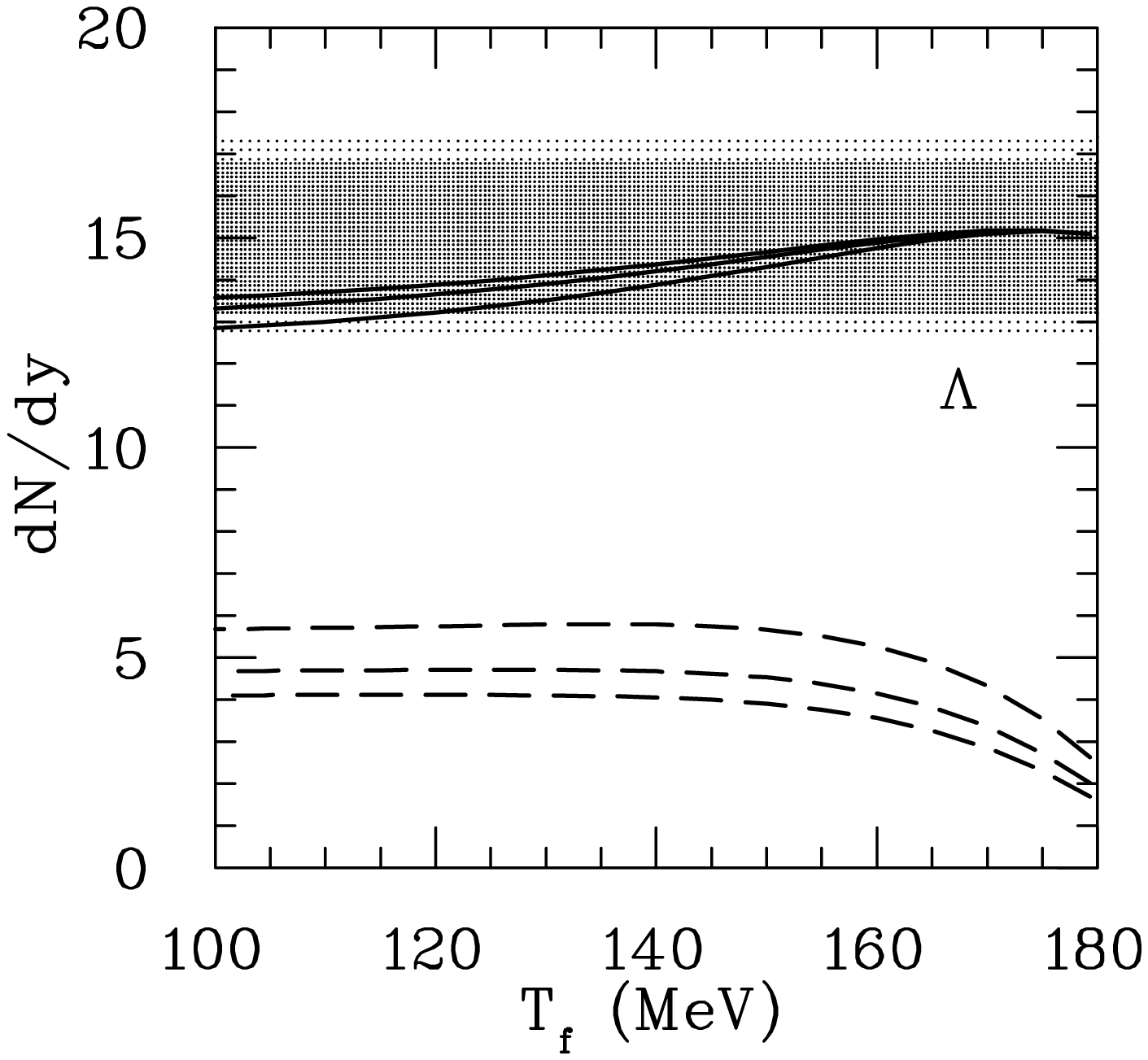,width=14cm}
\caption{The number of lambdas at central rapidity in 5\% most central
gold-gold collisions as a function of local temperature.  The solid
curves start with baryons in equilibrium, the lower set with no
baryons.  The central curve in each set uses the coupling coefficients
given in the Tables, while the other two use values larger and smaller
by a factor of two. With the exception of $\Lambda$, baryons unstable
to strong or weak decays have been decayed. The data are from the PHENIX
collaboration~\protect\cite{PHENIX} for the same centrality.
The dark band represents statistical and the light band systematic errors.}
\label{lambda_fig1}
\end{figure}

\begin{figure}
\epsfig{file=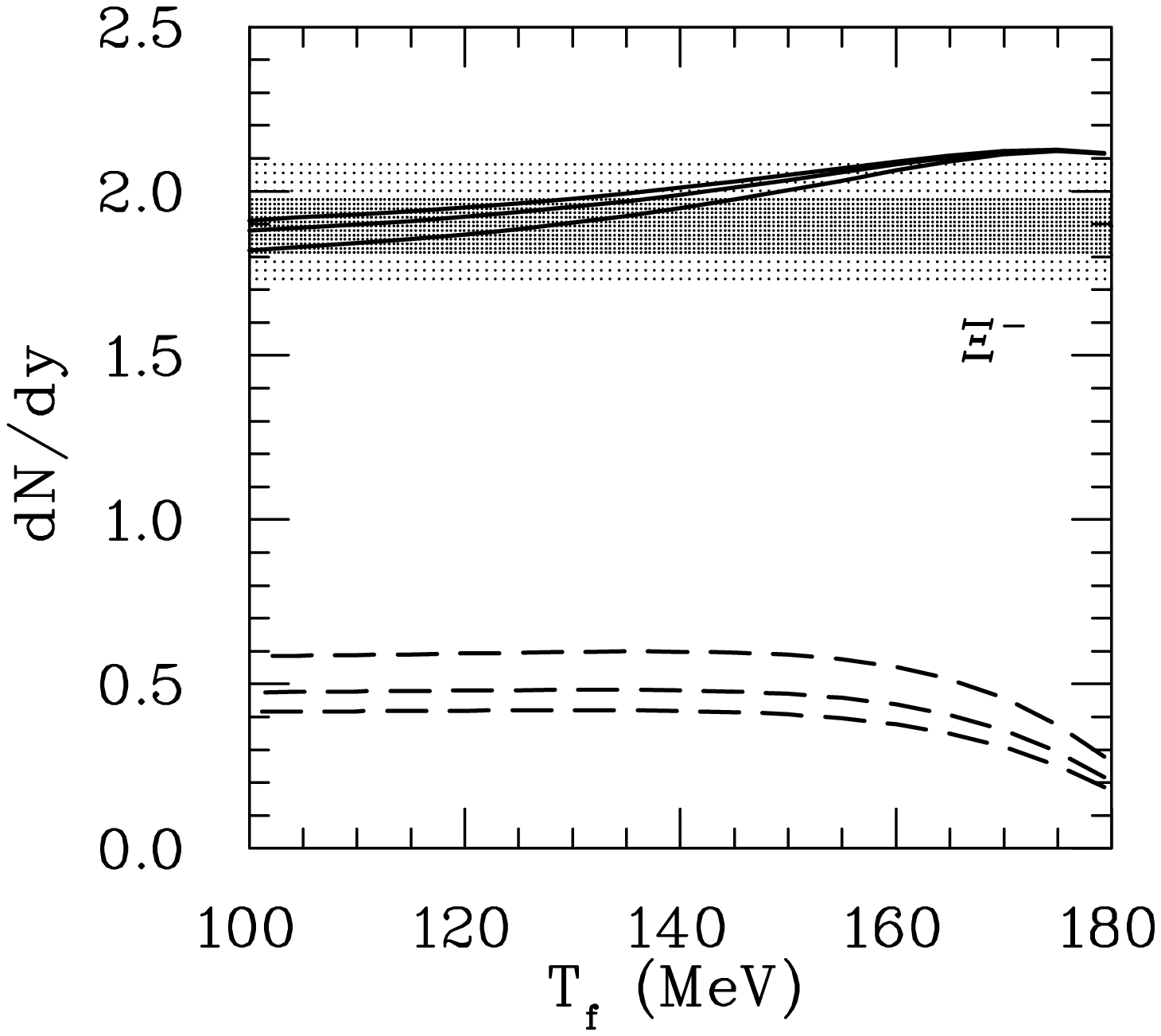,width=14cm}
\caption{The number of cascades at central rapidity in 5\% most central
gold-gold collisions as a function of local temperature.  The solid curves start
with baryons in equilibrium, the lower set with no baryons.  The central
curve in each set uses the coupling coefficients given in the Tables, while
the other two use values larger and smaller by a factor of two.  Baryons
unstable to strong decays have been decayed. The data are from the
STAR collaboration~\protect\cite{STAR} for 10\% most central collisions.
The dark band represents statistical and the light band systematic errors.}
\label{xi_fig1.ps}
\end{figure}

\begin{figure}
\epsfig{file=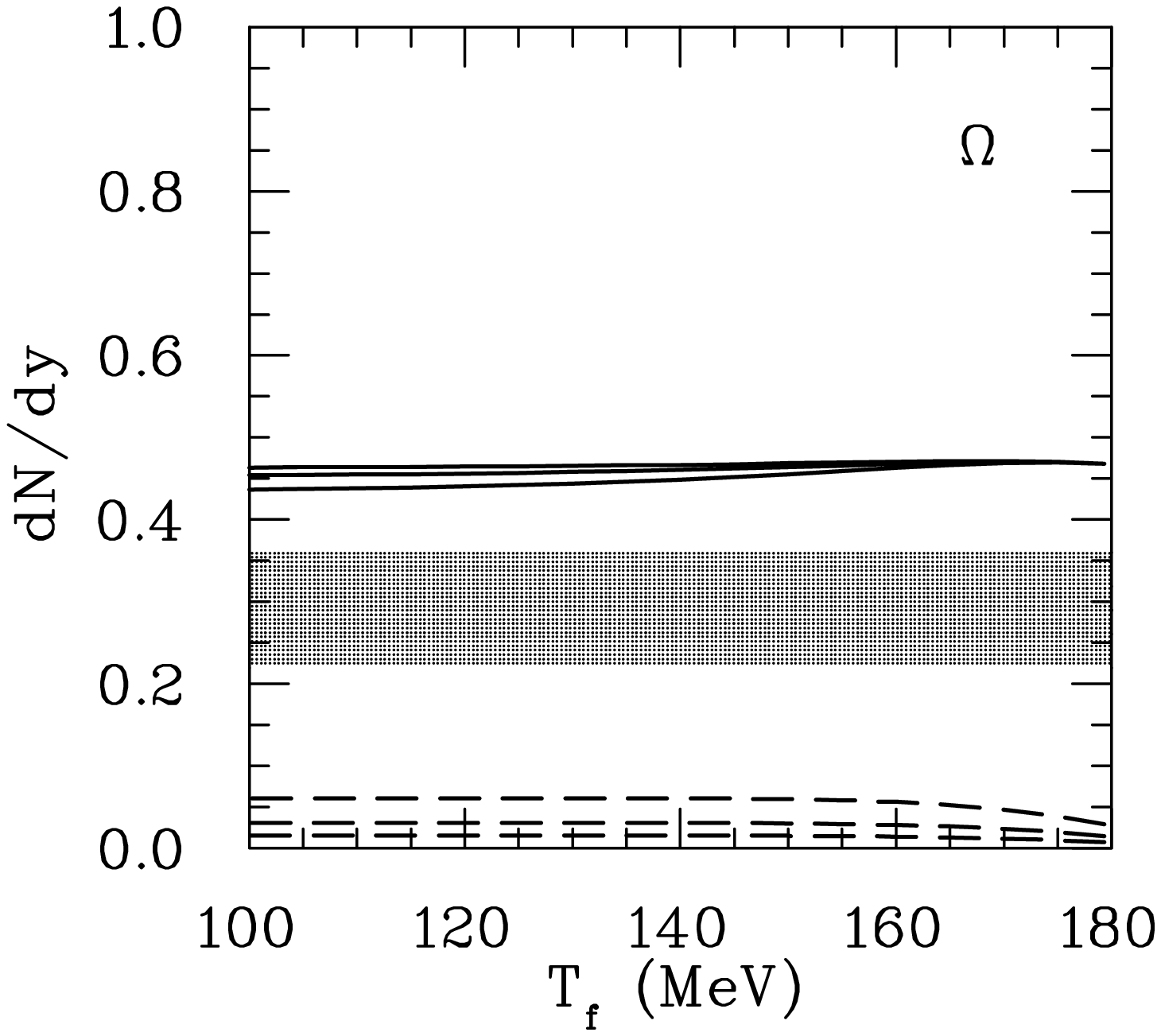,width=14cm}
\caption{The number of omegas at central rapidity in 5\% most central gold-gold
collisions as a function of local temperature.  The solid curves start
with baryons in equilibrium, the lower set with no baryons.  The central
curve in each set uses the coupling coefficients given in the Tables, while
the other two use values larger and smaller by a factor of two.  Baryons
unstable to strong decays have been decayed. The data are from the
STAR collaboration~\protect\cite{STAR} for 10\% most central collisions.
The dark band represents statistical errors only; the systematic errors
are smaller than the statistical ones.}
\label{omega_fig1.ps}
\end{figure}

\begin{figure}
\epsfig{file=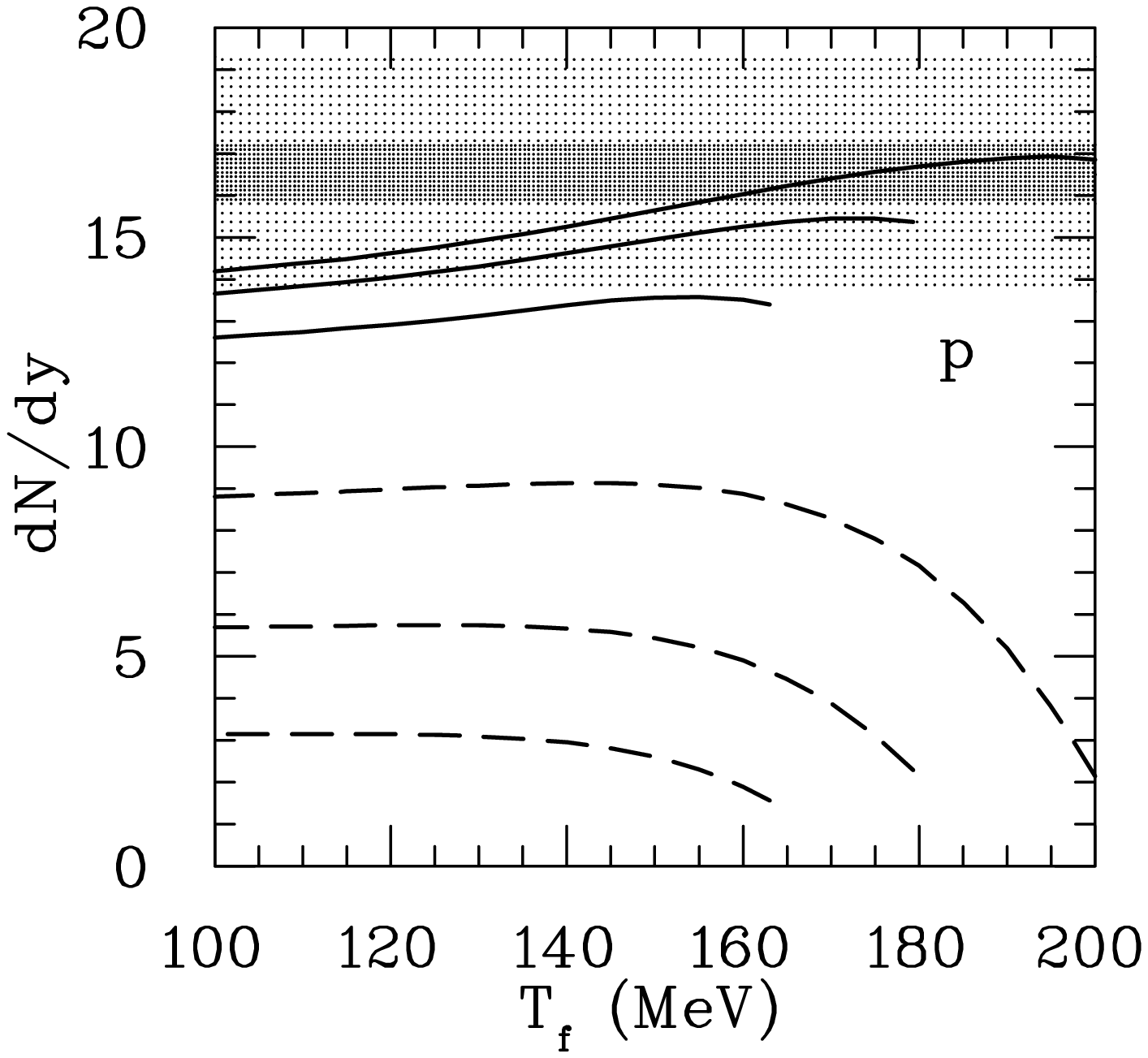,width=14cm}
\caption{Same as figure 2 except that the calculations are done for
three critical temperatures of 165, 180, and 200 MeV.  The coefficients
from the Tables are used.}
\label{proton_fig2}
\end{figure}

\begin{figure}
\epsfig{file=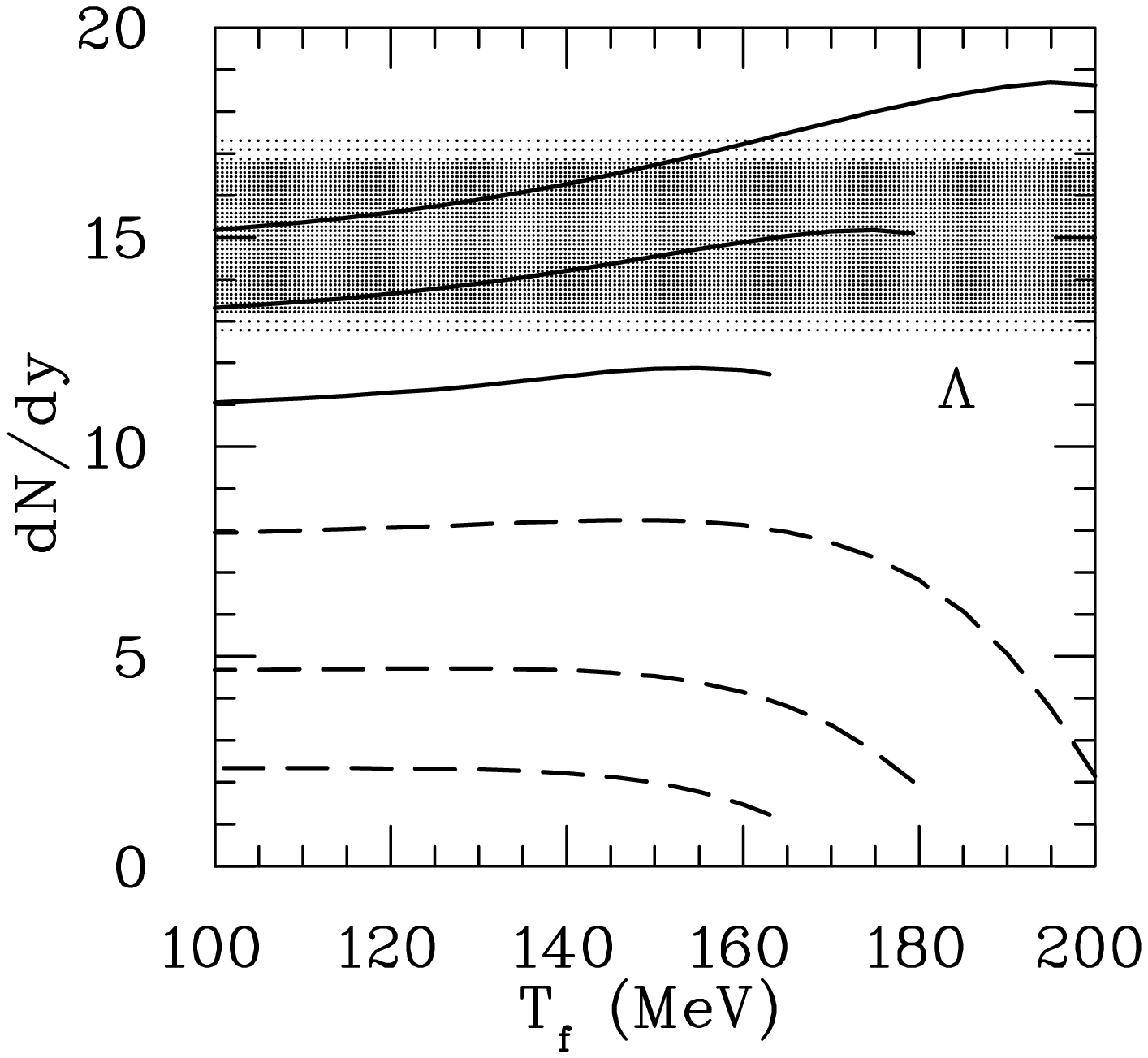,width=14cm}
\caption{ Same as figure 3 except that the calculations are done for
three critical temperatures of 165, 180, and 200 MeV.  The coefficients
from the Tables are used.}
\label{lambda_fig2}
\end{figure}

\begin{figure}
\epsfig{file=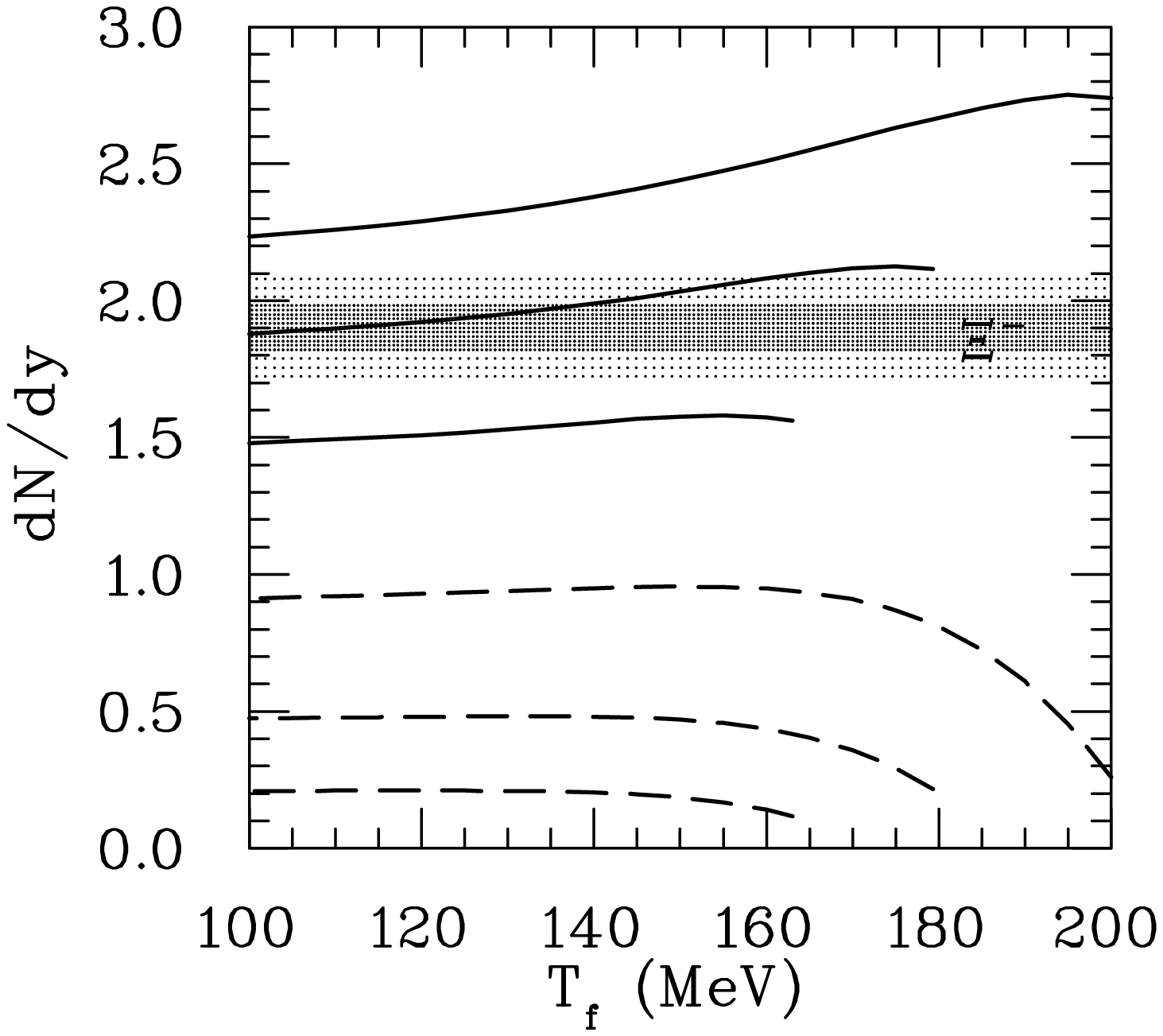,width=14cm}
\caption{ Same as figure 4 except that the calculations are done for
three critical temperatures of 165, 180, and 200 MeV.  The coefficients
from the Tables are used.}
\label{xi_fig2.ps}
\end{figure}

\begin{figure}
\epsfig{file=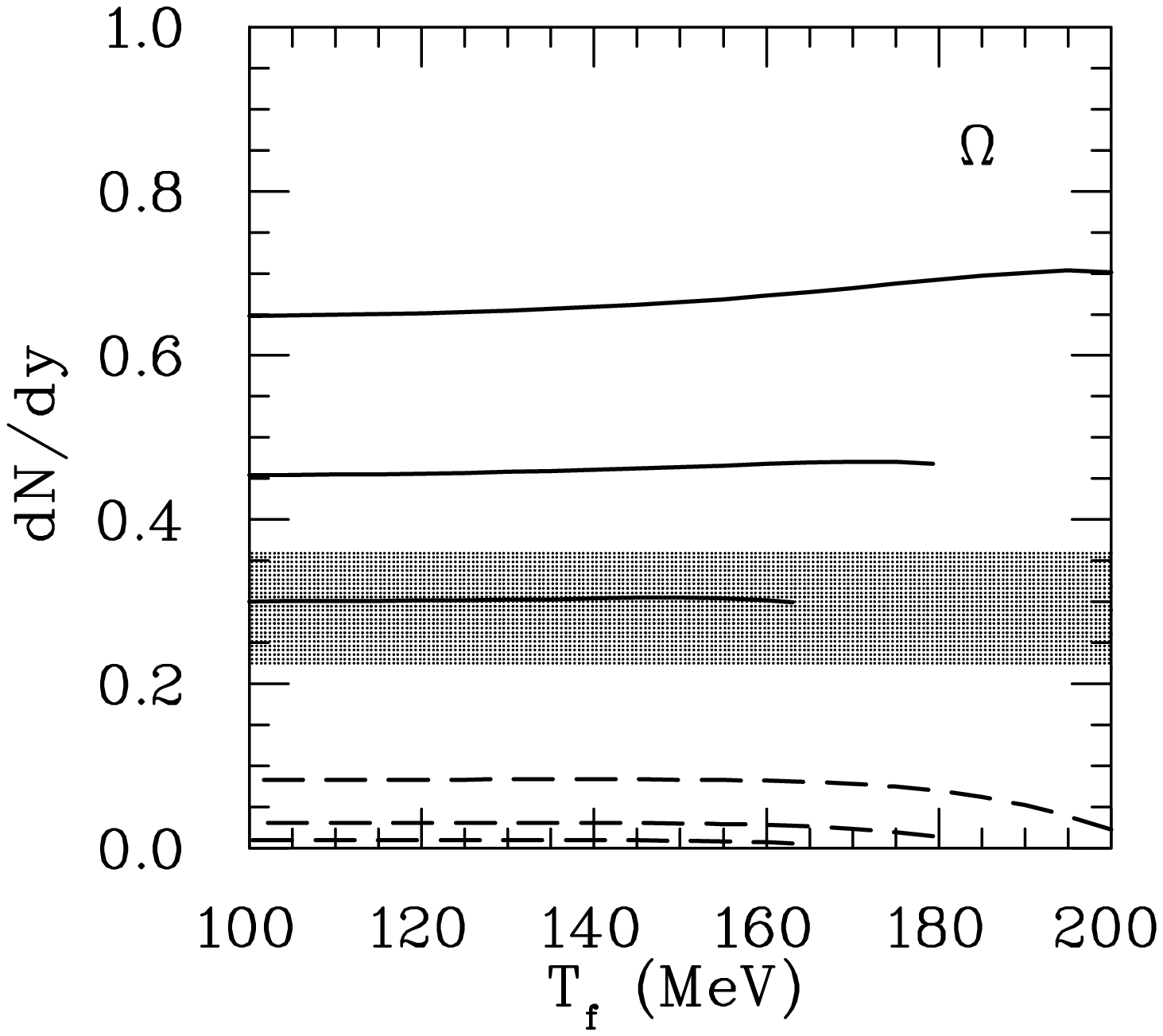,width=14cm}
\caption{ Same as figure 5 except that the calculations are done for
three critical temperatures of 165, 180, and 200 MeV.  The coefficients
from the Tables are used.}
\label{omega_fig2.ps}
\end{figure}


\begin{thebibliography} {99}

\bibitem{chem1} P. Braun-Munzinger, D. Magestro, K. Redlich, and J. Stachel
Phys. Lett. B {\bf 518}, 41 (2001).

\bibitem{chem2} J. Rafelski, J. Letessier, and G. Torrieri, Phys. Rev. C
{\bf 64}, 054907 (2001); Erratum-ibid. {\bf 65}, 069902 (2002).

\bibitem{phenix-pions} K. Adcox {\it et al.}  [PHENIX Collaboration],
Phys. Rev. Lett.  {\bf 88}, 242301 (2002).

\bibitem{QM2002}  Proceedings of Quark Matter 2002, Nantes, France, July 2002, 
 Nucl.\ Phys.\ A {\bf 715} (2003).

\bibitem{KS} J. Kapusta and I. Shovkovy, Phys. Rev. C {\bf 68}, 014901 (2003).

\bibitem{PDG}  Particle Data Group, D. E. Groom {\it et al.}, 
Eur. Phys. J. C {\bf 15}, 1 (2000).

\bibitem{hydrorev} P. Huovinen, arXiv:nucl-th/0305064;
P. F. Kolb and U. Heinz, arXiv:nucl-th/0305084.

\bibitem{hydro}
P. F. Kolb, P. Huovinen, U. W. Heinz and H. Heiselberg,
Phys. Lett. B {\bf 500}, 232 (2001);
P. F. Kolb, J. Sollfrank and U. W. Heinz,
Phys. Rev. C {\bf 62}, 054909 (2000).

\bibitem{PHENIX} K. Adcox {\it et al.} [PHENIX Collaboration],
Phys. Rev. Lett. {\bf 89}, 092302 (2002).

\bibitem{STAR} J. Adams {\it et al.} [STAR Collaboration],
arXiv:nucl-ex/0307024.


\end{thebibliography}
\end{document}